\documentclass[prd,twocolumn,superscriptaddress,altaffilletter,showpacs]{revtex4}
\usepackage{amssymb,amsmath}
\usepackage{comment}
\usepackage{graphicx}
\DeclareUnicodeCharacter{2212}{\textendash}

\begin{document}

\title{Observational redshift from general spherically symmetric black holes}
\author{Diego A. Martínez-Valera}
\email{diego.martinezval@alumno.buap.mx}
\affiliation{Facultad de Ciencias F\'{i}sico Matem\'{a}ticas, Benem\'{e}rita Universidad Aut\'{o}noma de Puebla, Ciudad Universitaria, Puebla, CP 72570, Puebla, Mexico}
\author{Mehrab Momennia}
\email{mmomennia@ifuap.buap.mx, momennia1988@gmail.com}
\affiliation{Instituto de F\'{\i}sica, Benem\'erita Universidad Aut\'onoma de Puebla,\\
Apartado Postal J-48, 72570, Puebla, Puebla, Mexico}
\affiliation{Instituto de F\'{\i}sica y Matem\'{a}ticas, Universidad Michoacana de San
Nicol\'as de Hidalgo,\\
Edificio C--3, Ciudad Universitaria, CP 58040, Morelia, Michoac\'{a}n, Mexico}
\author{Alfredo Herrera-Aguilar}
\email{aherrera@ifuap.buap.mx}
\affiliation{Instituto de F\'{\i}sica, Benem\'erita Universidad Aut\'onoma de Puebla,\\
Apartado Postal J-48, 72570, Puebla, Puebla, Mexico}
\date{\today }

\begin{abstract}
In this work, we obtain an expression for the total observational frequency shift of photons emitted by massive geodesic particles circularly orbiting a black hole in a general spherically symmetric background. Our general relations are presented in terms of the metric components and their derivatives that characterize the black hole parameters. As a concrete example of this general relativistic approach, a special case is studied by applying the formalism to a nonsingular black hole conformally related to the Schwarzchild solution that possesses a length scale parameter $l$ and an integer parameter $N$ in addition to the black hole mass. Besides, we express the nonsingular black hole mass in terms of the observational redshift/blueshift. Finally, we investigate the effects of the free parameters of the conformal gravity theory on the observational frequency shift and compare results with those of the standard Schwarzschild black hole. \vskip3mm

\end{abstract}

\maketitle



\section{Introduction}

Black holes are one of the most interesting compact objects obtained from Einstein’s general relativity theory and they are simple objects, completely described by only a handful of parameters, namely the black hole mass $M$ and angular momentum $J$. The existence of black holes has been identified to date through the pieces of evidence provided by observations of the orbital motion of stars at the center of our galaxy \cite{Ghez 1,Ghez 2, Genzel 1,Genzel 2}, captured shadow images of supermassive black holes hosted at the core of M87 and Milky Way galaxies \cite{EHT 1, EHT 2}, and gravitational wave detections by LIGO-Virgo collaborations \cite{LIGO 1, LIGO 2}. 

Recent discoveries in identifying black holes in nature have motivated astrophysicists to invent and develop new methods in order to obtain black hole parameters from different types of observations and gain a better understanding of the underlying physics. In this direction, a robust general relativistic method has been provided in \cite{HN-1,HN-2,KdS} which consists of obtaining the black hole parameters, such as mass and angular momentum, in terms of the observational redshift/blueshift of photons emitted by geodesic particles circularly orbiting in the vicinity of the Kerr black hole. Such a procedure allows one to obtain analytical expressions for the black hole's mass and spin in terms of directly measurable quantities \cite{HN-2}.

The primary work \cite{HN-1} has been introduced based on the kinematic frequency shift which is not a directly observable quantity and has been applied to a wide variety of black hole spacetimes, such as regular black holes in a spherically symmetric background \cite{Regular BH}, Kerr-Newmann black holes in de Sitter geometry \cite{Kerr-Newmann-dS}, and higher dimensional black holes \cite{High dim}. Then, this general relativistic method has been extended to incorporate observational total frequency shift in the mathematical modeling of black hole rotation curves \cite{HN-2}. The new extended method has been employed to express the parameters of static polymerized black holes in terms of the total redshift \cite{FuZhang}. More recently, it was shown that by employing a similar procedure, it is possible to extract the Hubble law from the Kerr black hole in asymptotically de Sitter spacetime and add a new independent general relativistic approach to measure the late-time Hubble constant \cite{KdS}.
 
 In addition, one can apply this general relativistic formalism to real astrophysical systems with available data on the frequency shift and positions of orbiting particles. This method has been used so far to estimate the mass-to-distance ratio of 17 supermassive black holes in the center of active galactic nuclei (AGNs) which possess an accretion disk of water vapor clouds \cite{ApJLNucamendi, Villalobos 1, Villaraos, Villalobos 2}. It was demonstrated that this relativistic approach allows us to quantify the gravitational redshift, hence identifying the relativistic effects in such astrophysical systems. Recently, by introducing the ``redshift rapidity" which is an observable element, we disentangled $M$ and $D$ in the Schwarzschild black hole spacetime and expressed mass and distance to the black hole just in terms of observational frequency shifts \cite{Schw BH}. The redshift rapidity has also been employed to express the mass, charge, and distance to the Reissner-Nordstr\"{o}m black hole in terms of a few directly observable quantities, such as the total frequency shift and aperture angle of the telescope \cite{RNBH}.

On the other hand, by considering recent electromagnetic and gravitational wave detections from black holes, we find that there are still large uncertainties in estimating the background test fields and the black hole parameters, such as electric charge and angular momentum \cite{Ghez 1,Ghez 2,Genzel 1,Genzel 2, EHT 1, EHT 2, LIGO 1, LIGO 2}. This uncertainty in obtaining parameters leads to the opportunity to explore their strong field regime and allows for the existence of black hole solutions beyond Einstein theory of relativity. The new black hole solutions of gravitational models beyond Einstein theory of relativity, like modified theories of gravitation and loop quantum gravity, as well as Einstein gravity coupled to various matter fields will contain new free parameters (characterizing the gravitational model) in addition to the black hole mass and angular momentum. The signature of these new free parameters describing the underlying modified gravity is encoded in the frequency shift of photons that we measure here on the Earth. Therefore, extending the general relativistic formalism \cite{HN-1, HN-2, KdS} to general spherically symmetric black hole spacetimes allows us to decode the information of the free parameters of new black hole solutions in modified theories of gravitation and probe new physics beyond Einstein gravity with the help of observational redshift.

In the present study, we aim to extend the general relativistic formalism \cite{HN-1, HN-2, KdS} to general spherically symmetric black hole spacetimes with the flat asymptote that would help to estimate or constrain the free parameters of the extended theories of gravitation. Then, we explore regular black holes in conformal gravity as a special example to demonstrate the application of this extension.

This paper is outlined as follows. In the next section, we briefly introduce the general spherically symmetric and static black hole background in asymptotically flat spacetime. In section \ref{Geodesics}, we consider the geodesic motion of the massive and massless particles in this background and express the nonvanishing components of the 4-velocity of the geodesic massive particles in terms of the metric functions that characterize the black hole parameters. Then, we continue to find the light bending parameter in terms of the metric functions with the help of the geodesic motion of emitted photons.
In section \ref{FrequencyShift}, we obtain a general expression for the total observational frequency shift of photons emitted by test particles circularly orbiting black holes in general spherically symmetric backgrounds in terms of the metric components by using the results obtained in Sec. \ref{Geodesics}. In Sec. \ref{CBH}, we employ the relations obtained in Secs. \ref{Geodesics}-\ref{FrequencyShift} in order to analyze the effects of free parameters of nonsingular black holes in conformal gravity on the observational frequency shift, hence showing an application of our general results. Finally, some concluding remarks are presented.

\section{Black hole spacetime configuration}
\label{BH configuration}

Here, we take into account a general static and spherically symmetric black
hole background with a flat asymptote. The line element takes the following
form in the Schwarzschild coordinates $\left( t,r,\theta ,\varphi \right) $
[we use $c=1=G$ units] 
\begin{equation}
ds^{2}=g_{\mu \nu }dx^{\mu }dx^{\nu }=g_{tt}dt^{2}+g_{rr}dr^{2}+g_{\theta
\theta }d\theta ^{2}+g_{\varphi \varphi }d\varphi ^{2},  \label{metric}
\end{equation}%
where the metric components $g_{tt}\left( r\right) $, $g_{rr}\left( r\right) 
$, and $g_{\theta \theta }\left( r\right) $ are arbitrary functions of the
radial coordinate $r$ as well as $g_{\varphi \varphi }\left( r,\theta
\right) =g_{\theta \theta }(r)\,sin^{2}\theta $. Hereafter, we remove the
arguments $r$ and $\theta $ from the metric functions for notational
simplicity \cite{CommentOnMetric}.

Since we are interested in asymptotically flat spacetimes, the metric
components of the line element (\ref{metric}) are required to behave as 
\begin{equation}
\lim_{r\rightarrow \infty }g_{tt}=-1,\ \ \lim_{r\rightarrow \infty
}g_{rr}^{-1}=1,\ \ \lim_{r\rightarrow \infty }g_{\theta \theta }=r^{2},
\end{equation}
at spatial infinity. In this paper, we are interested in investigating the
effects of spacetime curvature, produced by a general static and spherically
symmetric black hole described by the metric (\ref{metric}), on the
frequency shift of photons emitted by geodesic massive particles orbiting
the black hole. For instance, the photon sources could be stars, gas or water
masers in the case of AGNs.\newline

\section{Geodesics of timelike and null particles}

\label{Geodesics} The trajectory of particles revolving around the
spherically symmetric background (\ref{metric}) is governed by the following
equation of motion 
\begin{equation}
\frac{ds^2}{d\lambda^2}=g_{\mu\nu}\dot{x}^{\mu}\dot{x}^{\nu}=\kappa,
\label{geod}
\end{equation}
for both timelike ($\kappa=-1$) and null ($\kappa=0$) particles. Besides, $%
\dot{x}^{\mu}=dx^{\mu}/d\lambda$ denotes the derivative of the particle's
position where $\lambda$ is the affine parameter for photons and it
represents the proper time $\lambda=\tau$ in the case of massive particles.

Moreover, the background spacetime (\ref{metric}) possesses two Killing
vector fields as follows 
\begin{equation}
\xi^{\mu} = (1,0,0,0)\,\,\,\,\,\, \text{timelike Killing vector field},
\label{xi}
\end{equation}
\begin{equation}
\psi^{\mu}=(0,0,0,1) \,\,\,\,\,\,\text{rotational Killing vector field},
\label{psi}
\end{equation}
which characterize the symmetries of the spacetime and can be employed to
obtain the conserved quantities of the geodesic motion of the test particles
orbiting the black hole.

Due to the spherical symmetry of the line element (\ref{metric}), we
restrict the particle's motion to the equatorial plane $\theta =\pi /2$
without loss of generality. For inclined orbits of geodesic massive
particles as seen from the Earth, one can apply the rotation to the frame by
Euler angles. However, in the case of the accretion disks which can be only
observed from an edge-on view from the Earth \cite{edgeOn,edgeOnView,UGC3789,MCPXI}, this
equatorial plane has a great interest.

\subsection{Geodesics of massive particles}

The geodesic equation (\ref{geod}) for massive particles reads 
\begin{equation}
g_{\mu \nu}U^{\mu}U^{\nu}=-1,  \label{norm}
\end{equation}
where $U^{\mu}$ is the 4-velocity of the particle. On the other hand, the
conserved quantities associated with the Killing vector fields (\ref{xi})
and (\ref{psi}) are given by 
\begin{equation}
E = \frac{\Bar{E}}{m} = -\xi_{\mu}U^{\mu}=-g_{tt}U^t,  \label{CQE}
\end{equation}
\begin{equation}
L = \frac{\Bar{L}}{m} = \psi_{\mu}U^{\mu}=g_{\varphi\varphi}U^{\varphi},
\label{CQL}
\end{equation}
where $E$ and $L$ are, respectively, the total energy and angular momentum
of the test particle per unit mass. Now, by considering the particle motion confined in the equatorial plane where $U^\theta =0$, and substituting these relations
into Eq. (\ref{norm}), we have 
\begin{equation}
-\frac{1}{2}g_{tt}g_{rr}(U^r)^2 -\frac{g_{tt}}{2} -\frac{g_{tt}}{%
2g_{\varphi\varphi}}L^2=\frac{E^2}{2},  \label{Kin+Pot=E}
\end{equation}
which has the energy conservation law structure such that the first term is
the kinetic energy of the particle moving in an effective potential with the
following form 
\begin{equation}
V_{eff}=-\frac{1}{2}g_{tt}\left( 1+\frac{L^2}{g_{\varphi\varphi}}\right).
\label{Eff Pot}
\end{equation}

From Eq. (\ref{Kin+Pot=E}), the radial velocity of the massive particle
reads 
\begin{equation}
g_{rr}(U^{r})^{2}=-1-\frac{L^{2}}{g_{\varphi \varphi }}-\frac{E^{2}}{g_{tt}}%
\equiv V_{r}(r),  \label{RadVel}
\end{equation}%
that is a function of the radial coordinate $r$ only.

Taking into account the real astrophysical systems and spherical symmetry of the spacetime background, investigating circular equatorial motion is relevant. Besides, taking $\theta =\pi /2$ allows us
to express some relations in a simpler way without loss of generality. The
following conditions govern the massive particles in circular orbits 
\begin{equation}
V_{r}(r)=-1-\frac{E^{2}}{g_{tt}}-\frac{L^{2}}{g_{\varphi \varphi }}=0,
\end{equation}%
\begin{equation}
V_{r}^{\prime }(r)=\frac{E^{2}}{g_{tt}^{2}}g_{tt}^{\prime }+\frac{L^{2}}{%
g_{\varphi \varphi }^{2}}g_{\varphi \varphi }^{\prime }=0,
\end{equation}%
where the prime symbol denotes the derivative with respect to the radial
coordinate $r$. From these conditions, one can find $E$ and $L$ in terms of
the metric functions $g_{tt}$ and $g_{\varphi \varphi }$ and their
derivatives as below 
\begin{equation}
E^{2}=g_{\varphi \varphi }^{\prime }\left( \frac{g_{tt}^{2}}{g_{\varphi
\varphi }g_{tt}^{\prime }-g_{tt}g_{\varphi \varphi }^{\prime }}\right) ,
\label{Energ}
\end{equation}%
\begin{equation}
L^{2}=g_{tt}^{\prime }\left( \frac{g_{\varphi \varphi }^{2}}{%
g_{tt}g_{\varphi \varphi }^{\prime }-g_{\varphi \varphi }g_{tt}^{\prime }}%
\right) .  \label{AngMom}
\end{equation}

Now, with the aid of Eqs. (\ref{CQE}) and (\ref{CQL}), $t-$ and $\varphi -$
components of 4-velocity can be expressed as follows 
\begin{equation}
U_{e}^{t}=-\frac{E}{g_{tt}}=\sqrt{\frac{g_{\varphi \varphi }^{\prime }}{%
g_{\varphi \varphi }g_{tt}^{\prime }-g_{tt}g_{\varphi \varphi }^{\prime }}},
\label{Ut}
\end{equation}%
\begin{equation}
U_{e}^{\varphi }=\frac{L}{g_{\varphi \varphi }}=\sqrt{\frac{g_{tt}^{\prime }%
}{g_{tt}g_{\varphi \varphi }^{\prime }-g_{\varphi \varphi }g_{tt}^{\prime }}}%
,  \label{Uphi}
\end{equation}%
where subscript $e$ refers to the emitter radius $r_{e}$.

For stability of the circular orbits, it is required for the second
derivative of $V_{r}(r)$ to satisfy the following condition 
\begin{eqnarray}
&&V_{r}^{\prime \prime }(r)=-\frac{2E^{2}}{g_{tt}^{3}}\left( g_{tt}^{\prime
}\right) ^{2}+\frac{E^{2}}{g_{tt}^{2}}g_{tt}^{\prime \prime }  \notag \\
&&-\frac{2L^{2}}{g_{\varphi \varphi }^{3}}\left( g_{\varphi \varphi
}^{\prime }\right) ^{2}+\frac{L^{2}}{g_{\varphi \varphi }^{2}}g_{\varphi
\varphi }^{\prime \prime }\leq 0,  \label{Stability}
\end{eqnarray}%
where the equality indicates the innermost stable circular orbit (ISCO) and
characterizes the inner edge of the accretion disk.

It is possible to find an explicit expression for the second derivative of $%
V_{r}(r)$ in terms of the metric functions, with the aid of Eqs. (\ref{Energ}%
) and (\ref{AngMom}), which reads 
\begin{eqnarray}
V_{r}^{\prime \prime }(r) &=&\frac{1}{g_{tt}g_{\varphi \varphi }^{\prime }-g_{\varphi \varphi }g_{tt}^{\prime
}}\left\{ g_{\varphi \varphi }^{\prime }%
\left[ \frac{2}{g_{tt}}\left( g_{tt}^{\prime }\right) ^{2}-g_{tt}^{\prime
\prime }\right] \right.  \notag \\
&&\left. -g_{tt}^{\prime }\left[ \frac{2}{g_{\varphi \varphi }}\left(
g_{\varphi \varphi }^{\prime }\right) ^{2}-g_{\varphi \varphi }^{\prime
\prime }\right] \right\} .  \label{SecondDerivative}
\end{eqnarray}

\subsection{Geodesics of null particles}

In the case of massless particles, which follow null geodesics, the equation
of motion (\ref{geod}) reads 
\begin{equation}
g_{\mu \nu }k^{\mu }k^{\nu }=0,  \label{GeodPhot}
\end{equation}%
where $k^{\mu }=(k^{t},k^{r},k^{\theta },k^{\varphi })$ is the 4-wave vector
of the photons. 
These photons move outside the event horizon described by the metric (\ref%
{metric}), which depends on the explicit form of the functions $g_{tt}$, $%
g_{rr}$, and $g_{\varphi \varphi }$. By considering the motion of photons on the
equatorial plane, the component $k^{\theta }$ of the 4-wave vector vanishes,
hence Eq. (\ref{GeodPhot}) reduces to 
\begin{equation}
g_{tt}(k^{t})^{2}+g_{rr}(k^{r})^{2}+g_{\varphi \varphi }(k^{\varphi })^{2}=0.
\label{w4vect}
\end{equation}

Due to the symmetries of the spacetime described by the Killing vector
fields (\ref{xi})-(\ref{psi}), the following conserved quantities also hold
for the photon motion 
\begin{equation}
E_{\gamma }=-\xi _{\mu }k^{\mu }=-g_{tt}k^{t},
\end{equation}%
\begin{equation}
L_{\gamma }=\psi _{\mu }k^{\mu }=g_{\varphi \varphi }k^{\varphi },
\end{equation}
where $E_{\gamma }$ and $L_{\gamma }$ are the energy and angular momentum of the photons, respectively.

Now, by substituting these relations into the equation of motion of photons (%
\ref{w4vect}), we get 
\begin{equation}
g_{rr}(k^{r})^{2}+\frac{E_{\gamma }^{2}}{g_{tt}}+\frac{L_{\gamma }^{2}}{%
g_{\varphi \varphi }}=0.  \label{gEL}
\end{equation}

Note that for the points where the radial component of the 4-wave vector $k^{r}$
vanishes (which are the diametrically opposite points whose joint line is
perpendicular to the line of sight), Eq. (\ref{gEL}) simplifies in the
following way 
\begin{equation}
\frac{E_{\gamma }^{2}}{g_{tt}}+\frac{L_{\gamma }^{2}}{g_{\varphi \varphi }}%
=0.  \label{gEL2}
\end{equation}

By introducing the following definition of the light bending parameter in
the latter equation 
\begin{equation}
b_{\gamma }=\frac{L_{\gamma }}{E_{\gamma }},  \label{deflectPar}
\end{equation}%
we can find this parameter in terms of the properties of the spacetime
as follows 
\begin{equation}
b_{\gamma _{\pm }}=\pm \sqrt{-\frac{g_{\varphi \varphi }}{g_{tt}}}.
\label{deflectPar2}
\end{equation}

We recall that this relation shows the deflection of light sources when they
are located on either side of the black hole at the midline characterized by 
$\pm $\ sign.

\section{Frequency shift}

\label{FrequencyShift}

\begin{figure*}[tbh]
\centering
\includegraphics[width=0.4\textwidth]{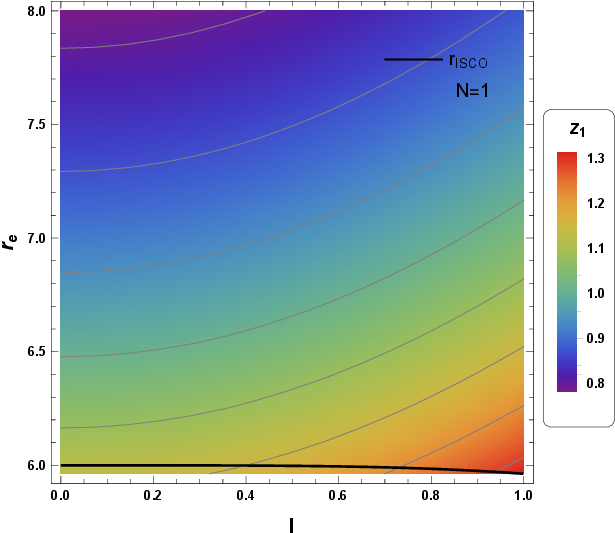} \includegraphics[width=0.4%
\textwidth]{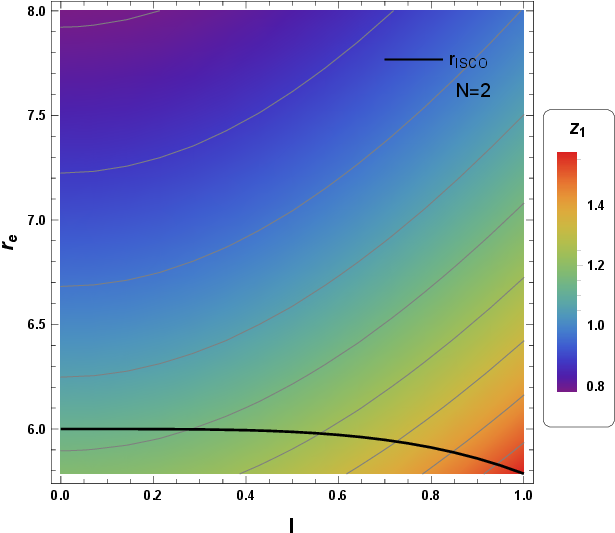}
\caption{The redshift in the $r_e-l$ plane for $M=1 $. The gray
curves stand for constant values of $z_1$ and the
black curve denotes $r_e=r_{ISCO}$.}
\label{RedDP}
\end{figure*}
\begin{figure*}[tbh]
\centering
\includegraphics[width=0.4\textwidth]{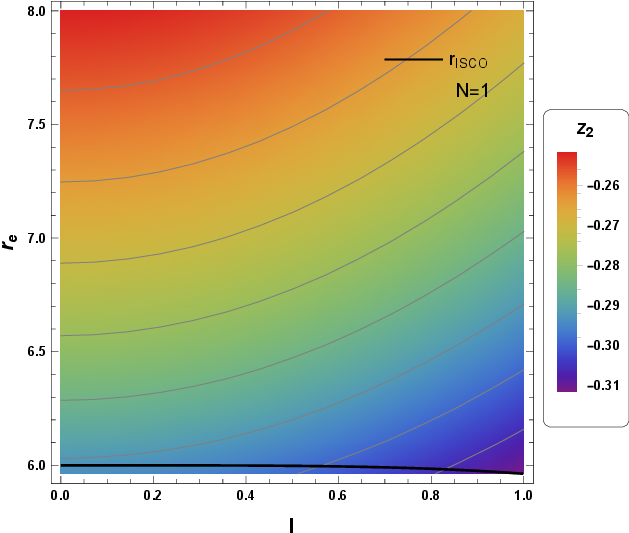} \includegraphics[width=0.4%
\textwidth]{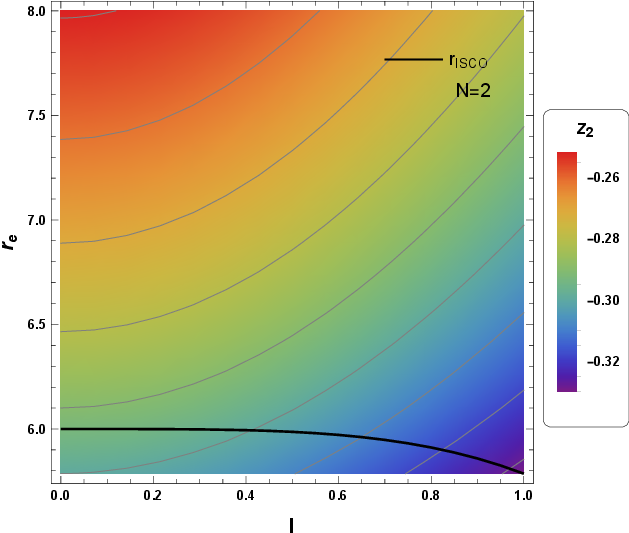}
\caption{The blueshift in the $r_e-l$ plane
for $M=1$. The gray curves
represent the constant levels of the blueshift and the black curve stands for $r_e=r_{ISCO}$.}
\label{BlueDP}
\end{figure*}
\begin{figure*}[tbp]
\centering
\includegraphics[width=0.29\textwidth]{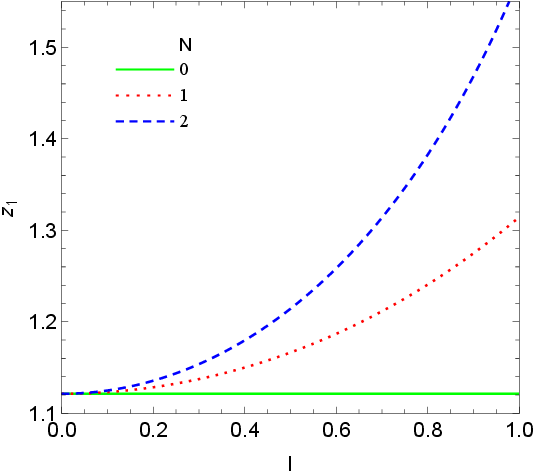} %
\includegraphics[width=0.303\textwidth]{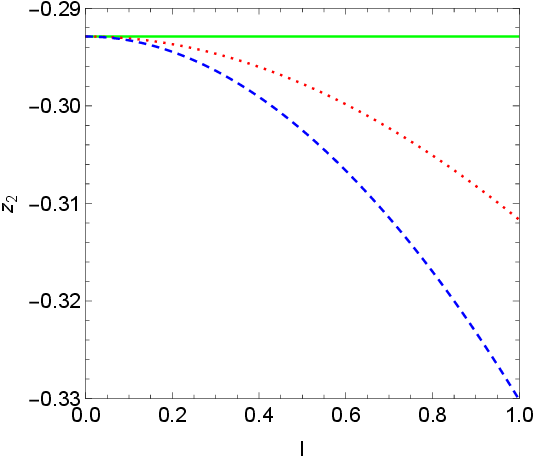} %
\includegraphics[width=0.3\textwidth]{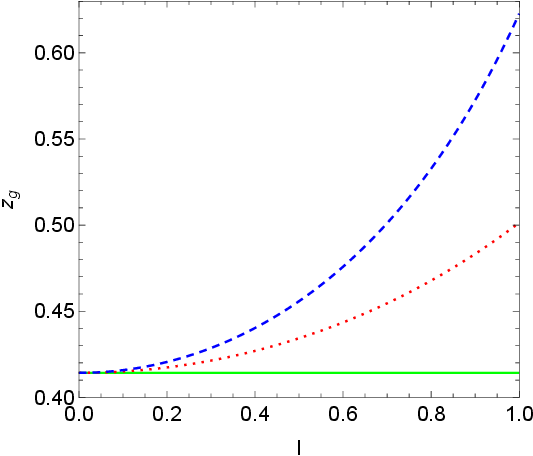}
\caption{The redshift $z_1$ (left), blueshift $z_2$ (middle)
and gravitational redshift $z_g$ (right) versus the length scale $l$ for
various values of $N$ and $M=1$. The emitter is orbiting circularly at 
$r_e = r_{ISCO}$ and the continuous green curves refer to the frequency shift in the Schwarzschild spacetime where $N=0$.}
\label{RBGfig}
\end{figure*}

The frequency of photons at some position $x_{p}^{\mu }=\left(
x^{t},x^{r},x^{\theta },x^{\varphi }\right) \mid _{p}$\ reads%
\begin{equation}
\omega _{p}=-\left( k_{\mu }U^{\mu }\right) \mid _{p}\,,  \label{freq}
\end{equation}%
where the index $p$ refers to either the point of emission $x_{e}^{\mu }$ or
detection $x_{d}^{\mu }$ of the photon.

The most general expression for shifts in the frequency $\omega _{p}$ in
static and spherically symmetric backgrounds of the form (\ref{metric}) can
be written as \cite{HN-1,HN-2,KdS}
\begin{eqnarray}
1 &+&z_{_{BH}}\!=\frac{\omega _{e}}{\omega _{d}}  \notag \\
&=&\frac{(E_{\gamma }U^{t}-L_{\gamma }U^{\varphi
}-g_{rr}U^{r}k^{r}-g_{\theta \theta }U^{\theta }k^{\theta })\mid _{e}}{%
(E_{\gamma }U^{t}-L_{\gamma }U^{\varphi }-g_{rr}U^{r}k^{r}-g_{\theta \theta
}U^{\theta }k^{\theta })\mid _{d}}\,.  \label{GeneralShift}
\end{eqnarray}

In addition, since the motion of the massive test particles is restricted to
the equatorial plane, implying that $U^{\theta }=0$, and from the circular
orbit condition $U^{r}=0$, the frequency shift formula (\ref{GeneralShift})
reduces to 
\begin{equation}
1+z_{_{BH}}\!=\frac{(E_{\gamma }U^{t}-L_{\gamma }U^{\varphi })\mid _{e}}{%
(E_{\gamma }U^{t}-L_{\gamma }U^{\varphi })\mid _{d}}\,.  \label{shift1}
\end{equation}

By considering the observer at a very large distance from the black hole,
its 4-velocity is given by $U_{d}^{\mu }=(1,0,0,0)$, and therefore the
frequency shift simplifies to 
\begin{equation}
1+z_{_{BH_{1,2}}}\!=U_{e}^{t}-b_{\gamma _{\mp }}U_{e}^{\varphi },  \label{shift2}
\end{equation}
where we used the light deflection parameter defined in Eq. (\ref{deflectPar}%
). 

One may note that $U_{e}^{\varphi }$ presented in Eq. (\ref{shift1})
also posses $\pm $ sings according to Eqs. (\ref{AngMom}) and (\ref{Uphi}) which amounts to
the rotation direction of massive particles ($+$/$-$ for counterclockwise/clockwise). However, due to the spherical symmetry of the background, it does not matter whether the particle is co-rotating or counter-rotating, hence we choose the plus
sign without loss of generality. 

Finally, by substituting Eqs. (\ref{Ut}), (%
\ref{Uphi}) and (\ref{deflectPar2}) into Eq. (\ref{shift2}), one obtains the
following general expression 
\begin{equation}
1+z_{_{BH_{1,2}}}=\frac{1}{\sqrt{g_{\varphi \varphi }g_{tt}^{\prime
}-g_{tt}g_{\varphi \varphi }^{\prime }}}\left( \sqrt{g_{\varphi \varphi
}^{\prime }}\pm \sqrt{\frac{g_{\varphi \varphi }}{g_{tt}}g_{tt}^{\prime }}%
\right) ,  \label{shift3}
\end{equation}%
for the shift in the frequency of photon sources circularly orbiting a
general spherically symmetric black hole spacetime at the radius $r=r_{e}$. In this relation, $%
z_{_{BH_{1}}}/z_{_{BH_{2}}}$ corresponds to the $+/-$ sign that represents
the redshift/blueshift at the midline and on either side of the line of
sight. It is important to remark here that the aforementioned expression is
the total frequency shift where the first term corresponds to the
gravitational frequency shift and the second term is the kinematic frequency
shift such that
\begin{equation}
z_{_{BH_{1,2}}}\!=z_{g}+z_{kin_{\pm }}.  \label{gPlusKin}
\end{equation}

\section{Black holes in conformal gravity as a special example}

\label{Examples} Here, in order to show the application of the results obtained in the
previous sections regarding the frequency shift of test particles revolving
spherically symmetric spacetimes, we take into account nonsingular black
holes in conformal gravity as a special example.

The curvature singularity at the center of black holes is a crucial and
outstanding problem and we need a quantum version of gravity to describe the
origin. Since there is no self-consistent and complete quantum gravity
theory so far, one can extend the gravitational theory beyond general
relativity in order to cure this essential singularity.

In this context, conformal gravity is an extended theory of gravitation that
provides a way to get rid of singularities by means of a conformal
transformation applied to the Schwarzschild metric $g_{\mu \nu }$, giving
rise to a new metric $\hat{g}_{\mu \nu }$ which is conformally related to $%
g_{\mu \nu }$ by the conformal factor $\Omega(x)$ as follows 
\begin{equation}
\hat{g}_{\mu \nu }\left( x\right) =\Omega \left( x\right) \,g_{\mu \nu
}\left( x\right) ,
\end{equation}%
where $\Omega \left( x\right) $ is a function of the spacetime coordinates
that satisfies 
\begin{equation}
0<\Omega \left( x\right) <\infty ,\,\,\,\,\Omega ^{-1}\left( x\right) \neq 0.
\end{equation}%

In this regard, one assigns the conformal symmetry to the Schwarzschild
background, hence the spacetime singularity can be removed after a suitable
conformal transformation. Nonsingular black holes conformally related to the
Schwarzschild solutions and the Kerr solutions have been proposed in \cite%
{FQG,Bambi} in which their spacetime is geodesically complete. In this
paper, we employ the following conformal factor \cite{Bambi} 
\begin{equation}
\Omega (r)=\left( 1-\frac{l^{2}}{r^{2}}\right) ^{2N},
\label{ConformalFactor}
\end{equation}%
where $l$ is a length factor and $N$ is a positive integer. Therefore, the new
line element $d\hat{s}^{2}$ conformally related to the Schwarzschild black
holes $ds_{Schw}^{2}=g_{\mu \nu }dx^{\mu }dx^{\nu }$\ with the conformal
factor $\Omega (r)$\ is given by%
\begin{equation}
d\hat{s}^{2}=\Omega (r)ds_{Schw}^{2}=\Omega (r)g_{\mu \nu }dx^{\mu }dx^{\nu
}=\hat{g}_{\mu \nu }dx^{\mu }dx^{\nu },  \label{ConformalLE}
\end{equation}%
where the Schwarzschild metric components are as follows%
\begin{equation}
g_{\mu \nu }=diag\left[ -\left( 1-\frac{2M}{r}\right) ,\left( 1-\frac{2M}{r}%
\right) ^{-1},r^{2},r^{2}\,sin^{2}\theta \right] ,  \label{SchwMetric}
\end{equation}%
in which $M$\ is the total mass of the Schwarzschild black hole. Now, by
taking into account Eqs. (\ref{ConformalFactor})-(\ref{SchwMetric}), the
explicit form of the metric components of the conformal nonsingular black
holes reads%
\begin{equation}
\hat{g}_{tt}(r)=-\left( 1-\frac{l^{2}}{r^{2}}\right) ^{2N}\left( 1-\frac{2M}{%
r}\right) ,  \label{Cgtt}
\end{equation}%
\begin{equation}
\hat{g}_{rr}(r)=\left( 1-\frac{l^{2}}{r^{2}}\right) ^{2N}\left( 1-\frac{2M}{r%
}\right) ^{-1},  \label{Cgrr}
\end{equation}%
\begin{equation}
\hat{g}_{\theta \theta }(r)=\left( 1-\frac{l^{2}}{r^{2}}\right) ^{2N}r^{2},
\label{Cgtheta}
\end{equation}%
\begin{equation}
\hat{g}_{\varphi \varphi }(r,\theta )=\hat{g}_{\theta \theta }(r)sin^{2}\theta .
\label{Cgphi}
\end{equation}

As the next step, we can consider massive geodesic test particles circularly
orbiting the conformal black hole described by (\ref{ConformalLE}) and apply
all the formalism developed previously. Besides, by substituting the
expressions for the metric components (\ref{Cgtt})-(\ref{Cgphi}) into Eq. (%
\ref{SecondDerivative}), $r_{ISCO}$ can be obtained by solving 
\begin{eqnarray}
&&r^{5}-6Mr^{4}+2l^{2}(N-1)r^{3}+  \notag \\
&&\frac{4l^{2}}{M}\left[ Nl^{2}\left( 2N-1\right) -3M^2(N-1)\right] r^{2}-  \notag \\
&&l^{4}\left( 2N-1\right) \left[ (20N+1)r-6M(4N+1)\right] =0,  \label{rISCO}
\end{eqnarray}%
and we have stable orbits for the radii $r\geq r_{ISCO}$. Note that this
equation is a $5$th order polynomial which cannot be solved analytically and
we shall use its numerical solutions in our future analysis. A numerical investigation of the solutions to Eq. (\ref{rISCO}) in black hole mass unit for $N=1,2$ and $0\leq l\leq 1$ indicates that there is only one positive definite root larger than the photon sphere radius $r_{ph}$ and the rest of the
roots are either imaginary/complex-valued or smaller than $r_{ph}$ [$r_{ph}$ can be obtained through the condition $\left( g_{tt}^{\prime }g_{\varphi \varphi }\right) _{r=r_{ph}}=\left(
g_{tt}g_{\varphi \varphi }^{\prime }\right) _{r=r_{ph}}$ and this constraint is deduced from Eq. (\ref{gEL}) by considering the conditions $k^r=0$ and $(k^r)^{\prime}=0$]. Hence, the positive definite root of Eq. (\ref{rISCO}) satisfying $r_{ISCO}>r_{ph}$ indicates the ISCO radius.

However, the black curves in Figs. \ref{RedDP}-\ref{BlueDP} show the dependency of $r_{ISCO}$ on the free parameters $N$ and $l$ of the theory. From these figures, we see that for nonvanishing values of the free parameters, $r_{ISCO}^{(CG)}$ in conformal gravity is always less than $r_{ISCO}^{(Schw)}=6$ for the Schwarzschild black hole case, $r_{ISCO}^{(CG)}<r_{ISCO}^{(Schw)}$. It is also important to remark that although it is possible to have circular orbits
for $r_e<r_{ISCO}$, these orbits are not stable. Hence, the region of interest corresponds to values of $r$ greater than $r_{ISCO}$ for which bounded orbits are stable.

On the other hand, from Eqs. (\ref{shift3}) and (\ref{Cgtt})-(\ref{Cgphi}), the total frequency shift in the
conformal black hole background reads%
\begin{eqnarray}
1 &+&z_{_{1,2}}=\frac{1}{\sqrt{\left( 1-\frac{l^{2}}{r_e^{2}}\right)
^{2N}(r_e-3M)(r_e-2M)(r_e^{2}-l^{2})}}\times   \notag \\
&&\left( \sqrt{r_e(2Nl^{2}+r_e^{2}-l^{2})(r_e-2M)}\right.   \notag \\
&&\pm \left. \sqrt{r_e\left[ Mr_e^{2}+2Nl^{2}r_e-Ml^{2}\left( 4N+1\right) \right] }%
\right) ,  \label{TotalShift}
\end{eqnarray}%
where one can identify 
\begin{equation}
1+z_{g}=\frac{\sqrt{r_e(2Nl^{2}+r_e^{2}-l^{2})}}{\sqrt{\left( 1-\frac{l^{2}}{r_e^{2}}%
\right) ^{2N}(r_e-3M)(r_e^{2}-l^{2})}},  \label{GravShift}
\end{equation}%
and 
\begin{equation}
z_{kin_{\pm }}=\pm \frac{\sqrt{r_e\left[ Mr_e^{2}+2Nl^{2}r_e-Ml^{2}\left(
4N+1\right) \right] }}{\sqrt{\left( 1-\frac{l^{2}}{r_e^{2}}\right)
^{2N}(r_e-3M)(r_e-2M)(r_e^{2}-l^{2})}},  \label{KinShift}
\end{equation}%
as, respectively, the gravitational frequency shift and kinematic frequency
shift satisfying $z_{_{1,2}}=z_{g}+z_{kin_{\pm }}$. Note that these formulas
reduce to the Schwarzschild black hole case \cite{ApJLNucamendi} for either $%
N=0$\ or $l=0$, as it should be. In addition, since $z_{1,2}$ are observable quantities, the aforementioned relations
can be employed alongside the real astrophysical data in order to estimate
(or constraint) the free parameters $N\ $and $l$\ appearing in this model of
conformal gravity.

Note that the length scale parameter $l$ has been constrained with the help of x-ray observational data of the supermassive black hole in 1H0707-495 to be $l\leq 0.6$ for $N=2$ \cite{CGTest1} and $l\leq0.225$ for $N=1$ \cite{CGTest2}. Therefore, we concentrate our attention on the restricted values $N=1,2$ and $0\leq l \leq 1$ for the free parameters of the conformal gravity in our analysis. Figs. \ref{RedDP}-\ref{BlueDP} illustrate the density plots of the redshift and blueshift in the $r_e-l$ parameter space for $N=1,2$. As one can see from these figures, $r_e$ and $l$ have opposite effects on the redshift/blueshift and as $r_e$ ($l$) increases, the shift in frequency decreases (increases). Therefore, it is expected that these parameters balance the frequency shift at some points, denoted by the continuous gray curves in Figs. \ref{RedDP} and \ref{BlueDP}, which represent constant level sets for $z_{1,2}$.

In addition, from Fig. \ref{RBGfig}, one can see that the redshift, blueshift, and gravitational redshift increase when the free parameter $N$ increases. Moreover, this figure shows that as the parameter $N$ takes greater values, the frequency shift curves increase faster as a function of $l$ since $N$ appears as an exponent in the conformal factor $\Omega$. As the free parameters take higher values, $z_{1,2}$ significantly deviates from the Schwarzschild redshift/blueshift depicted in Fig. \ref{RBGfig} as $N=0$. It is worth mentioning that although $z_g$ is positive and increases as the free parameters increase (see the right panel of Fig. \ref{RBGfig}), $z_2$ is still negative and acquires lower values for nonvanishing $N$ and $l$ since $z_{kin_{-}}$-term in Eq. (\ref{TotalShift}) is dominant (see the middle panel of Fig. \ref{RBGfig}).

Furthermore, by defining $R=1+z_{1}$ and $B=1+z_{2}$, one can readily verify
\begin{equation}
RB=\frac{1}{\Omega (r_{e})\left( 1-\frac{2M}{r_{e}}\right) },
\label{RBproduct}
\end{equation}%
which leads to an expression for the mass of the nonsingular black hole in terms of the
observable frequency shifts $z_{1}$ and $z_{2}$ as well as the free
parameters $l$ and $N$ of the conformal gravity theory in the following way 
\begin{eqnarray}
M &=&\frac{RB-\left( \frac{r_{e}^{2}}{r_{e}^{2}-l^{2}}\right) ^{2N}}{2RB}%
r_{e}  \notag \\
&=&\frac{(1+z_{1})(1+z_{2})-\left( \frac{r_{e}^{2}}{r_{e}^{2}-l^{2}}\right)
^{2N}}{2(1+z_{1})(1+z_{2})}r_{e},  \label{BHmassRelation}
\end{eqnarray}%
where reduces to the Schwarzschild black hole mass formula for either
vanishing $N$\ or $l$, as it should be. It is worth mentioning that the maximum values of $z_{1}$ and $z_{2}$ for revolving particles in stable circular orbits occur at the radius $r_{ISCO}$. In this regard, the frequency shift values on the black curves in Figs. \ref{RedDP}-\ref{BlueDP} indicate the possible maximum values of $z_{1}$ and $z_{2}$ for each black hole configuration. In addition, figure \ref{RBGfig} is evaluated for the emitter orbiting circularly the black hole at $r_{e}=r_{ISCO}$, hence the curves indicate the maximum values of the frequency shift. In this picture, the frequency shift vanishes as the emitter radius tends to infinity.

From an observational point of view, we can refer to the measured values of $z_1$ and $z_2$ from H$_2$O megamaser systems on the accretion disks of supermassive black holes hosted at the core of AGNs. For instance, observations from the megamaser system at the center of NGC 4258 galaxy show that the redshift lies in the range $4\times10^{-3}<z_1<6\times10^{-3}$, the blueshift is within $-1.7\times10^{-3}<z_2<-1.0\times10^{-3}$, and the emitter radius $r_e$ extends in the sub-parsec region from $0.04 pc$  to
$0.5 pc$ from the center of the disk \cite{NGC4258}. Similarly, from the data set of the megamaser systems NGC 1194, NGC 2273, NGC 2960, NGC 6264, NGC 6323, and UGC 3789, we find the ranges $1.1\times10^{-3}<z_1<3.3\times10^{-3}$ for the redshift, $-2.6\times10^{-3}<z_2<-0.9\times10^{-3}$ for the blueshift, and $0.028\ pc<r_e<1.33\ pc$ for the emitter radius \cite{Villaraos,MCP1,MCP3}. One may note that applying our general relativistic formalism to such astrophysical systems, including the ones mentioned above, allows us to quantify the gravitational redshift $z_g$ (\ref{GravShift}), hence identifying a general relativistic effect in these systems \cite{ApJLNucamendi, Villalobos 1, Villaraos, Villalobos 2}. In the case of megamaser systems, it was shown that the gravitational redshift ranges $1.6\times10^{-7}<z_g<3.1\times10^{-5}$, depending on the mass of the supermassive black hole and its distance to the emitter.

In addition, it is worth noting that by taking into account the frequency
shift formulas (\ref{TotalShift}) and the conformal factor (\ref%
{ConformalFactor}) for the special case $N=1$, one can show that 
\begin{equation}
\mathcal{X}\equiv RB=\frac{r_{e}}{(r_{e}-2M)\Omega (r_{e})},
\end{equation}%
\begin{equation}
\mathcal{Y}\equiv \left( R+B\right) ^{2}=\frac{4\left[ 2-\Omega ^{1/2}(r_{e})%
\right] r_{e}}{(r_{e}-3M)\Omega ^{3/2}(r_{e})}.
\end{equation}

Now, by employing the trigonometric solution of cubic equations for the irreducible case \cite{Uspensky}, we solve these relations for $M$ and $l$ to find the nonsingular black hole mass and length
scale only in terms of the observational redshift/blueshift and orbital
parameter of the emitter as follows%
\begin{eqnarray}
    M= \frac{r_e}{2}\left(1-\frac{3\mathcal{Y}\,sec^2\left(\frac{p+\sigma}{3}\right)}{4\left(8\mathcal{X}+3\mathcal{Y}\right)}\right),
\end{eqnarray}
\begin{eqnarray}
    l_{_{\pm}}=r_e\left(1\pm\frac{2\sqrt{8\mathcal{X}+3\mathcal{Y}}\,cos\left(\frac{p+\sigma}{3}\right)}{\sqrt{3\mathcal{X Y}}}\right)^{\frac{1}{2}},
\end{eqnarray}
where the parameter $\sigma$ is given by
\begin{eqnarray}
    \sigma=arccos\left(-\frac{24\mathcal{X}\sqrt{3\mathcal{X}\mathcal{Y}}}{\left(8\mathcal{X}+3\mathcal{Y}\right)^{\frac{3}{2}}}\right),\label{paramter sigma}
\end{eqnarray}
and the solutions $M$ and $l_{_{+}}$ for $p=\pi$ are valid for $\frac{M}{r_e}<\frac{1}{6}$, whereas $M$ and $l_{_{-}}$ for vanishing $p=0$ are valid for $\frac{M}{r_e}\geq\frac{1}{6}$.

\label{CBH}

\section{Discussion and final remarks}

In this work, we have extended a general relativistic method \cite{HN-1,HN-2,KdS} for measuring the black hole parameters to general spherically symmetric spacetimes by considering massive geodesic particles circularly orbiting a static black hole. Then, we have analytically obtained a general formula for the observational frequency shift of photons emitted by the test particles in terms of the metric functions and their derivatives that characterize the black hole parameters. 

In addition, we have studied a special case by applying the former results to a nonsingular black hole conformally related to the Schwarzchild solutions as a concrete example of this general relativistic approach. The nonsingular black hole mass was expressed in terms of the observational redshift/blueshift. Furthermore, we have investigated the effects of the free parameters of the conformal gravity theory on the observational redshift and compared results with those of the standard Schwarzschild black hole. Specifically, we have seen that for nonvanishing values of the free parameters $N$ and $l$ of the conformal gravity theory, the total observational redshift/blueshift increases. 

Furthermore, it would be interesting to estimate (or constraint) the free parameters $N$ and $l$ with the help of real astrophysical data of supermassive black holes hosted at the core of AGNs by making use of Bayesian fitting methods and comparing results to the previous estimations based on x-ray data \cite{CGTest1,CGTest2}. We leave this study for future work.

Within this general relativistic formalism, the detector receives the information of spacetime background encoded in the frequency shift of photons emitted by the photon source. If the spacetime background is constructed based on gravitational theories beyond Einstein gravity, it would characterized by some new free parameters of the theory, hence the information of these free parameters encoded in the observational redshift. In this regard, if we develop the mathematical modeling of the general relativistic formalism \cite{HN-1,HN-2,KdS} to general spherically symmetric spacetimes, this would help us to extract the information of the free parameters of the underlying theory from observations. Therefore, the generalization of this formalism to general spherically symmetric spacetimes, that contain an arbitrary number of the free parameters and the metric functions depend on the modified gravity, is a crucial extension of this approach and could be useful to extract information on black hole parameters and test extended theories of gravity.


\section*{Acknowledgments}

All authors are grateful to CONACYT for support under Grant No.
CF-MG-2558591; M.M. also acknowledges SNI and was supported by CONACYT through the postdoctoral Grant No. 31155. A.H.-A. thanks
SNI and PROMEP-SEP, and was supported by Grant VIEP-BUAP No. 122.



\begin{thebibliography}{9}
\bibitem{CommentOnMetric} Although there is a more standard representation
of a general spherically symmetric line element with a flat asymptote, we
found it convenient to employ the given form (\ref{metric})\ for future
usage of the final results.
\end{thebibliography}
\end{document}